\documentclass[12pt]{article}
\usepackage{graphicx}
\usepackage{amssymb}
\usepackage{amsmath}

\begin{document}

\begin{titlepage}

\begin{flushright}
ICRR-Report-617-2012-6\\
IPMU~12-0139
\end{flushright}

\vskip 1.35cm

\begin{center}

{\large 
{\bf Stochastic Approach to Flat Direction during Inflation} 
}

\vskip 1.2cm

Masahiro Kawasaki$^{a,b}$
and
Tomohiro Takesako$^a$ \\

\vskip 0.4cm

{ \it$^a$Institute for Cosmic Ray Research,
University of Tokyo, Kashiwa 277-8582, Japan}\\
{\it $^b$Kavli Institute for the Physics and Mathematics of the Universe,
University of Tokyo, Kashiwa 277-8568, Japan}\\

\date{\today}

\begin{abstract} 
We revisit the time evolution of a flat and non-flat direction system during inflation.
In order to take into account quantum noises in the analysis,
we base on stochastic formalism and solve coupled Langevin equations numerically.
We focus on a class of models in which tree-level Hubble-induced mass is not generated.
Although the non-flat directions can block the growth of the flat direction's variance in principle,
the blocking effects are suppressed by the effective masses of the non-flat directions.
We find that the fate of the flat direction during inflation is determined by one-loop radiative corrections and non-renormalizable terms as usually considered,
if we remove the zero-point fluctuation from the noise terms.
\end{abstract}

\end{center}
\end{titlepage}

\section{Introduction}
\label{sec:intro}
In the inflationary universe, quantum fluctuations of a scalar field are extended to superhorizon scale by the cosmic expansion.
If the scalar field is massless, the variance of the superhorizon fluctuation grows as $(H/2\pi)^2 H t$, where $H$ is the Hubble parameter during inflation and $t$ is the cosmic time~\cite{Linde:1982uu}.
Inflaton, which is responsible for inflation, is one of the almost massless scalar field in the inflationary epoch and thus have quantum fluctuations.
The slow-roll equation of motion for an inflaton is, however, usually described as a classical one.
In order to analyze the effect of quantum fluctuations on the Inflaton dynamics, stochastic approach has been proposed~\cite{Starobinsky1984, Sasaki:1987gy}\footnote{
For applications to various inflation models,  for example see Ref.~\cite{Nakao:1988yi}}.
In the stochastic approach, one integrates all of the superhorizon modes, whose definition is somewhat artificial, to obtain so-called IR mode.
The equation of motion for IR mode is described by coupled Langevin equations with noise terms.
These noise terms represent the quantum "kicks" by horizon crossing modes and drive IR mode to evolve stochastically.

In the context of supersymmetric standard models, there are many flat directions in scalar field space~\cite{Gherghetta:1995dv}.
However, it is well-known that, in supergravity framework, an effective mass of the order of Hubble scale is generically generated to a flat direction during inflation~\cite{hep-ph/9405389, Dine:1995uk}.
Such a large effective mass can be avoided, if one assumes the D-term inflation scenario~\cite{hep-ph/9405389} or imposes a Heisenberg symmetry on K$\ddot {\text{a}}$hler potential~\cite{Gaillard:1995az}. 
Without tree-level Hubble-induced masses, one may have considered that one-loop radiative corrections  and non-renormalizable terms determine the time evolution of a flat direction. 
Recently, Ref.~\cite{Enqvist:2011pt} analyzed interacting systems which consist of flat and non-flat directions without the tree-level Hubble-induced mass terms using the stochastic approach.
In Ref.~\cite{Enqvist:2011pt}, it is insisted that the variance of the flat direction will saturate because of its effective mass in the potential arising from the couplings with the non-flat directions.
However, in Ref.~\cite{Enqvist:2011pt}, the noise terms do not reflect the time evolutions of the effective masses.
Thus, the conclusion in Ref.~\cite{Enqvist:2011pt} can be essentially modified when we properly include the effective mass effects in the noise terms.
In this study, we formulate the effective mass effects in the noise terms and then analyze the time evolution of a flat and non-flat direction system.

The construction of this paper is as following:
in section~\ref{sec:sto}, we review the formalism of stochastic approach to a single scalar field case. We also describe contributions of the zero-point fluctuation for the noise terms.
In section~\ref{sec:corr}, we formulate the noise terms including the effective mass of the scalar field.
Then, in section~\ref{sec:num}, we consider a concrete system and
show the results of the numerical calculation for coupled Langevin equations.
Section~\ref{sec:conc} is devoted to conclusion.

\section{Stochastic approach}
\label{sec:sto}
In this section, we briefly review the stochastic approach to the time evolution of a scalar field~\cite{Starobinsky1984, Sasaki:1987gy}.
In the following, we consider a real scalar field $\phi(x)$ during inflation
and take Friedmann-Robertson-Walker metric: $\mathrm{d} s^2 = \mathrm{d} t^2 - a^2(t) \mathrm{d} \bold{x}^2$, where $a (t)$ is the scale factor.
For convenience, we also formulate contributions from the zero-point fluctuation for noise terms, which we use in the following analysis.

\subsection{Langevin equation}
Let us start with the equation of motion for a real scalar field $\phi(x)$:
\begin{equation}\label{eq:eom}
\begin{split}
\ddot \phi + 3 H \dot \phi - \frac{\nabla^2}{a^2} \phi + \frac{\partial V (\phi)}{\partial \phi} = 0,
\end{split}
\end{equation}
where `` $\dot{}$ " represents the time derivative $\partial_t$ and $V (\phi)$ is a potential for $\phi(x)$.
We define the mode expansion for $\phi(x)$ as
\begin{equation}
\begin{split}
\phi (x) = \int \frac{\mathrm{d}^3 \bold{k}}{(2 \pi)^{3/2}} \left( a_{\bold{k}} \varphi_k(t) + a_{- \bold{k}} \varphi_k^*(t) \right) \mathrm{e}^{i \bold{k} \cdot \bold{x}},
\end{split}
\end{equation}
where the annihilation and creation operators $a_{\bold{k}}, a_{\bold{k}}^{\dagger}$ satisfy the following commutation relation:
\begin{equation}
\begin{split}
[a_{\bold{k}}, a_{\bold{k'}}^{\dagger}] = \delta^{(3)} (\bold{k} - \bold{k'}).
\end{split}
\end{equation}
Here and hereafter, $\bold{k}$ is a comoving momentum.
After linearizing the equation of motion Eq.~(\ref{eq:eom}), the mode function $\varphi_k (t)$ can be determined from the following equation:
\begin{equation}\label{eq:mode-eom}
\begin{split}
\ddot \varphi_k (t) + 3 H \dot \varphi_k (t) + \frac{k^2}{a^2} \varphi_k (t) + \tilde m^2 \varphi_k (t) = 0,
\end{split}
\end{equation}
where the effective mass $\tilde m$ is given by
\begin{equation}\label{eq:effave}
\begin{split}
\tilde m^2 = \left\langle \frac{\partial^2 V (\phi)}{\partial \phi^2} \right\rangle.
\end{split}
\end{equation}
Here, $\langle \cdots \rangle$ is the spatial average inside the horizon during inflation.
In the numerical calculation, we replace this spatial average by IR mode field value which is defined below.
The solution for Eq.~(\ref{eq:mode-eom}) is given by
\begin{equation}\label{eq:mode-func}
\begin{split}
\varphi_k (t)
&= \sqrt{\frac{\pi}{4 k^3}} H \mathrm{e}^{i \left( \frac{\pi \nu}{2} + \frac{\pi}{4} \right)} \left( \frac{k}{a H} \right)^{3/2} H_{\nu}^{(1)} \left( \frac{k}{a H} \right),
\end{split}
\end{equation}
where $ H_{\nu}^{(1)} (z)$ is a first kind Hankel function of order $\nu$ and we have chosen the Bunch-Davis vacuum~\cite{Bunch:1978yq}.
Here, $\nu$ is given by $\nu^2 = \frac{9}{4} - \frac{\tilde m^2}{H^2}$.
We note that this mode function~(\ref{eq:mode-func}) satisfies $\varphi_k (t) \to \frac{1}{\sqrt{2 k} a} \mathrm{e}^{i \frac{k}{a H}}$ in the short wave length limit $\frac{k}{a H} \to \infty$.

Now, let us define IR mode for the scalar field $\phi (x)$ and its conjugate $\pi (x) \equiv \dot \phi (x)$ as
\begin{equation}
\begin{split}
&\Phi (\bold{x}, t) 
= (\phi (\bold{x}, t))_{\text{IR}}
\equiv \int \frac{\mathrm{d}^3 k}{(2 \pi)^{3/2}}  \theta (\epsilon a (t) H - k) \left( a_{\bold{k}} \varphi_k(t) + a_{- \bold{k}} \varphi_k^*(t) \right) \mathrm{e}^{i \bold{k} \cdot \bold{x}}, \\
&\Pi (\bold{x}, t) 
= (\pi (\bold{x}, t))_{\text{IR}}
\equiv \int \frac{\mathrm{d}^3 k}{(2 \pi)^{3/2}}  \theta (\epsilon a (t) H - k) \left( a_{\bold{k}} \dot \varphi_k(t) + a_{- \bold{k}} \dot \varphi_k^*(t) \right) \mathrm{e}^{i \bold{k} \cdot \bold{x}},
\end{split}
\end{equation}
where $\theta (t)$ is the step function.
Then, Eq.~(\ref{eq:eom}) is reduced to the following coupled Langevin equations for IR modes~\cite{Sasaki:1987gy, Enqvist:2011pt}:
\begin{equation}\label{eq:Langevin}
\begin{split}
&\dot \Phi (\bold{x}, t) = \Pi (\bold{x}, t) + s^{(\phi)} (\bold{x}, t), \\
&\dot \Pi (\bold{x}, t) = - 3 H \Pi (\bold{x}, t) - \frac{\partial V (\Phi)}{\partial \Phi} + s^{(\pi)} (\bold{x}, t).
\end{split}
\end{equation}
Here the noise terms $s^{(\phi)} (\bold{x}, t), s^{(\pi)} (\bold{x}, t)$ are given by
\begin{equation}
\begin{split}
&s^{(\phi)} (\bold{x}, t) = \epsilon a(t) H^2 \int \frac{\mathrm{d}^3 k}{(2 \pi)^{3/2}}  \delta (\epsilon a (t) H - k) \left( a_{\bold{k}} \varphi_k(t) + a_{- \bold{k}} \varphi_k^*(t) \right) \mathrm{e}^{i \bold{k} \cdot \bold{x}}, \\
&s^{(\pi)} (\bold{x}, t) = \epsilon a(t) H^2 \int \frac{\mathrm{d}^3 k}{(2 \pi)^{3/2}}  \delta (\epsilon a (t) H - k) \left( a_{\bold{k}} \dot \varphi_k(t) + a_{- \bold{k}} \dot \varphi_k^*(t) \right) \mathrm{e}^{i \bold{k} \cdot \bold{x}}.
\end{split}
\end{equation}
Eq.~(\ref{eq:Langevin}) is what we would like to derive in this subsection and is the basis of this study.
Although we have considered a single real scalar field case, it is straightforward to derive Langevin equations for a multi-field system.
In Section~\ref{sec:num}, we analyze Langevin equations for a flat and non-flat direction system.

In the numerical calculation, we need correlation functions between the noise terms, which are integrated over a short interval $[t, t + \mathrm{d} t]$.
The integrated correlation functions have the following forms:
\begin{equation}
\begin{split}
S^{(\phi)} (r, t;\mathrm{d} t)
&\equiv \int_t^{t + \mathrm{d} t} \mathrm{d} t_1 ~\int_t^{t + \mathrm{d} t} \mathrm{d} t_2~\langle \text{vac}|~ s^{(\phi)} (\bold{x_1}, t_1) s^{(\phi)} (\bold{x_2}, t_2)~| \text{vac} \rangle \\
&= \left( \frac{H}{2 \pi} \right)^2 H \mathrm{d} t~ j_0 (\epsilon a H r)~\frac{\pi}{2} \epsilon^3 |H_{\nu}^{(1)} (\epsilon)|^2 \times 
\begin{cases}
&1~~(\nu = \text{real}), \\
&\mathrm{e}^{- \pi \mu}~~(\nu = i \mu),
\end{cases}
\end{split}
\end{equation}
\begin{equation}
\begin{split}
&S^{(\pi)} (r, t;\mathrm{d} t) \\
&\equiv \int_t^{t + \mathrm{d} t} \mathrm{d} t_1 ~\int_t^{t + \mathrm{d} t} \mathrm{d} t_2~\langle \text{vac}|~ s^{(\pi)} (\bold{x_1}, t_1) s^{(\pi)} (\bold{x_2}, t_2)~| \text{vac} \rangle \\
&=  \left( \frac{H^2}{2 \pi} \right)^2 H \mathrm{d} t~ j_0 (\epsilon a H r)~\frac{\pi}{2} \epsilon^3~\left| \left( \frac{3}{2} - \nu \right) H_{\nu}^{(1)} (\epsilon) + \epsilon H_{\nu - 1}^{(1)} (\epsilon) \right|^2  \times 
\begin{cases}
&1~~(\nu = \text{real}), \\
&\mathrm{e}^{- \pi \mu}~~(\nu = i \mu),
\end{cases}
\end{split}
\end{equation}
where $r = |\bold{x_1-x_2}|$ and $j_0 (x)$ is a spherical Bessel function of order $0$.
Here, we have divided the expressions according to the cases where $\nu$ is real and $\nu = i \mu~(\mu>0)$ is pure imaginary.

\subsection{The zero-point fluctuation}
In this subsection, we formulate contributions from the zero-point fluctuation for the noise correlation functions.
The mode function for the zero-point fluctuation is given by
\begin{equation}
\begin{split}
\varphi_k^{\text{zero}} (t) &= \frac{1}{\sqrt{2 \omega_k (t)} a(t)} \mathrm{e}^{- i \int^t \mathrm{d} t' \sqrt{k^2/a^2 + \tilde m^2}}, \\
\dot \varphi_k^{\text{zero}} (t) &= - \left[ 1 + \frac{1}{2} \frac{\tilde m^2 / H^2}{(k/aH)^2 + \tilde m^2 / H^2} + i \sqrt{ \left( \frac{k}{a H} \right)^2 + \frac{\tilde m^2}{H^2}} \right] H \varphi_k^{\text{zero}} (t).
\end{split}
\end{equation}
Here, we have defined $\omega_k (t) = \sqrt{k^2 + \tilde m^2 a^2(t)}$.
Then, we obtain the following noise correlation functions:
\begin{equation}\label{eq:zero-cor}
\begin{split}
\bar S^{(\phi) \text{zero}} (r, N;\mathrm{d} N) &= \frac{\mathrm{d} N}{(2 \pi)^2} j_0 (\epsilon a H r) \frac{\epsilon^3}{\sqrt{\epsilon^2 + \tilde m^2 / H^2}}, \\
\bar S^{(\pi) \text{zero}} (r, N;\mathrm{d} N) &= \frac{\mathrm{d} N}{(2 \pi)^2} j_0 (\epsilon a H r) \frac{\epsilon^3}{\sqrt{\epsilon^2 + \tilde m^2 / H^2}} \left[ \left( 1 + \frac{1}{2} \frac{\tilde m^2 / H^2}{\epsilon^2 + \tilde m^2 / H^2} \right)^2 + \epsilon^2 + \tilde m^2 / H^2 \right].
\end{split}
\end{equation}
Here, we have normalized the correlation functions by Hubble scale $H$ as $\bar{S}^{(\phi)} (r, N;\mathrm{d} N) = S^{(\phi)} (r, t;\mathrm{d} t)/H^2$, $\bar{S}^{(\pi)} (r, N;\mathrm{d} N) = S^{(\pi)} (r, t;\mathrm{d} t)/H^4$
and $N = H t$ is the e-folding number (we need dimensionless quantities for the following numerical analysis).
In particular, these expressions become simple for $\epsilon^2 \ll \tilde m^2 / H^2$:
\begin{equation}\label{eq:zeronoise}
\begin{split}
\bar S^{(\phi) \text{zero}} (r, N;\mathrm{d} N) &\simeq \frac{\mathrm{d} N}{(2 \pi)^2} j_0 (\epsilon a H r) \epsilon^3 \frac{H}{\tilde m}, \\
\bar S^{(\pi) \text{zero}} (r, N;\mathrm{d} N) &\simeq \frac{\mathrm{d} N}{(2 \pi)^2} j_0 (\epsilon a H r) \epsilon^3 \frac{\tilde m}{H} \left( 1 + \frac{9}{4 \tilde m^2 / H^2} \right),
\end{split}
\end{equation}
and for $\tilde m^2 / H^2 = 0 ~(\nu = 3/2)$:
\begin{equation}
\begin{split}
\bar S^{(\phi) \text{zero}} (r, N;\mathrm{d} N) &= \frac{\mathrm{d} N}{(2 \pi)^2} j_0 (\epsilon a H r) \epsilon^2, \\
\bar S^{(\pi) \text{zero}} (r, N;\mathrm{d} N) &= \frac{\mathrm{d} N}{(2 \pi)^2} j_0 (\epsilon a H r) \epsilon^2 \left( 1 + \epsilon^2 \right).
\end{split}
\end{equation}

Now we mention the treatment of the contributions from the zero-point fluctuation.
Although, as far as we know, we have no guiding principle to remove or to keep the zero-point fluctuation in the noise terms,
we remove the zero-point fluctuation basically.
The case where the zero-point fluctuation is left in the noise terms is discussed separately below.
Fortunately, however, as we see in the next section, the contributions from the zero-point fluctuation are sufficiently suppressed by powers of $\epsilon$. 
Thus, as long as we take sufficiently small $\epsilon$, the effect of the zero-point fluctuation in noise terms are negligible.
An exception is the case for large effective mass $\tilde m \gg H$, where the zero-point fluctuation dominates the noise terms.
In the next section, we discuss the effect of zero-point fluctuation on noise terms concretely.

\section{Correlation functions between noise terms}
\label{sec:corr}
In this section, we formulate the noise correlation functions which we use for the numerical analysis in the next section.
In particular, we take care of the effective mass dependences of the noise correlation functions. 
The noise correlation functions have quite different form between $\nu = \sqrt{\frac{9}{4} - \frac{\tilde m^2}{H^2}}$ ($- \infty < \frac{\tilde m^2}{H^2} \leq \frac{9}{4}$)
and $\nu = i \mu = i \sqrt{\frac{\tilde m^2}{H^2} - \frac{9}{4}}$ ($\frac{\tilde m^2}{H^2} \geq \frac{9}{4}$).
Thus, in this section, we formulate the noise correlation functions for each case.

\subsection{case : $\nu$ is real}
Here, we consider the case where $\nu = \sqrt{\frac{9}{4} - \frac{\tilde m^2}{H^2}}$ is real ($- \infty < \frac{\tilde m^2}{H^2} \leq \frac{9}{4}$).
From the approximation formulae Eq.~(\ref{eq:app-wellknown2}), (\ref{eq:app-int}) in Appendix, we obtain the following form of Hankel function for $\epsilon \ll 1$:
\begin{equation}
\begin{split}
|H_{\nu}^{(1)} (\epsilon)|^2
&\simeq 
\begin{cases}  
& \frac{2^{2 \nu} \Gamma (\nu)}{\pi^2} \epsilon^{-2 \nu}~~\left(- \infty < \frac{\tilde m^2}{H^2} < \frac{9}{4} \right), \\
&\frac{4}{\pi^2} \left( \ln \frac{\epsilon}{2} \right)^2~~\left( \frac{\tilde m^2}{H^2} \simeq \frac{9}{4} \right).
\end{cases}
\end{split}
\end{equation}
Then, the dimensionless noise correlation function is given by
\begin{equation}\label{eq:noise-cor}
\begin{split}
\bar{S}^{(\phi)} (r, N; \mathrm{d} N) 
&= \frac{\mathrm{d} N}{(2 \pi)^2}~j_0 (\epsilon a H r) \times \frac{\pi}{2} \epsilon^3~|H_{\nu}^{(1)} (\epsilon)|^2\\
&\simeq \frac{\mathrm{d} N}{(2 \pi)^2}~j_0 (\epsilon a H r) \times
\begin{cases}
&\frac{2^{2 \nu} \Gamma (\nu)^2}{2 \pi} \epsilon^{3 - 2 \nu}~~\left(- \infty < \frac{\tilde m^2}{H^2} \leq 2 \right), \\
& 2 \epsilon^2 \nu + \frac{2 \epsilon^3}{\pi} \left( \ln \frac{\epsilon}{2} \right)^2 \left( 1 - 2 \nu \right)~~\left(2 \leq \frac{\tilde m^2}{H^2} \leq \frac{9}{4} \right). \\
\end{cases}\\
\end{split}
\end{equation}
To make our analysis easy, we naively extrapolate $\frac{\tilde m^2}{H^2} = 2$ to $\frac{\tilde m^2}{H^2} = \frac{9}{4}$ in the second line in Eq.~(\ref{eq:noise-cor}).
We use this extrapolated formula for the numerical calculation in Section.~\ref{sec:num}.
There is no special reason for choosing $\frac{\tilde m^2}{H^2}  = 2$ as the boundary,
however, we have checked that the numerical results in Section.~\ref{sec:num} is insensitive to $\frac{\tilde m^2}{H^2} \geq 2$ for $\epsilon = 10^{-2}$.
Note that the zero-point contribution~Eq.~(\ref{eq:zero-cor}) is not a leading order in $\epsilon$ compared with Eq.~(\ref{eq:noise-cor}).
Thus, the treatment of the zero-point contribution does not matter in this case.

In the same way, we approximate the noise correlation functions for the conjugate $\pi (x)$ ($\epsilon \ll 1$) as
\begin{equation}\label{eq:noise-cor-pi}
\begin{split}
\bar{S}^{(\pi)} (r, N; \mathrm{d} N) 
&= \frac{\mathrm{d} N}{(2 \pi)^2}~j_0 (\epsilon a H r) \times \frac{\pi}{2} ~\epsilon^3~\left| \left( \frac{3}{2} - \nu \right) H_{\nu}^{(1)} (\epsilon) + \epsilon H_{\nu - 1}^{(1)} (\epsilon) \right|^2 \\
&\simeq \frac{\mathrm{d} N}{(2 \pi)^2}~j_0 (\epsilon a H r) \times 
\begin{cases}
&\frac{2^{2 \nu} \Gamma (\nu)^2}{2 \pi} \epsilon^{3 - 2 \nu} \left( \frac{3}{2} - \nu \right)^2~~\left(- \infty < \frac{\tilde m^2}{H^2} \leq 2 \right), \\
& 2 \epsilon^2  \nu + \frac{9 \epsilon^3}{2 \pi} \left( \ln \frac{\epsilon}{2} \right)^2 (1 - 2 \nu)~~\left( 2 \leq \frac{\tilde m^2}{H^2} \leq \frac{9}{4} \right).
\end{cases}
\end{split}
\end{equation}
In the second line in Eq.~(\ref{eq:noise-cor-pi}) for $- \infty < \frac{\tilde m^2}{H^2} \leq 2$, the expression becomes just $0$ for $\frac{\tilde m^2}{H^2} = 0$ $(\nu = \frac{3}{2})$.
With more accurate approximation, it should be $\frac{\mathrm{d} N}{(2\pi)^2} j_0 (\epsilon a H r) \times \frac{\pi \epsilon^3}{2} \epsilon^2 |H_{1/2}^{(1)} (\epsilon)|^2 \simeq \frac{\mathrm{d} N}{(2\pi)^2} j_0 (\epsilon a H r) \times \epsilon^4$.
Since this is sufficiently small for $\epsilon \ll 1$, we omit this $\epsilon^4$ contribution and use Eq.~(\ref{eq:noise-cor-pi}) in the following numerical calculation.
We note that the numerical results in Section.~\ref{sec:num} is insensitive to the boundary $\frac{\tilde m^2}{H^2} = 2$ for $\epsilon = 10^{-2}$.
Also, note that the zero-point contribution is not a leading order in powers of $\epsilon$ compared with Eq.~(\ref{eq:noise-cor-pi}) 
and thus the treatment of the zero-point contribution does not matter here.

\subsection{case : $\nu = i \mu, ~(\mu \geq 0)$}
Next, we consider the case where the effective mass is very large: $\nu = i \mu = i \sqrt{\frac{\tilde m^2}{H^2} - \frac{9}{4}}$ ($\frac{\tilde m^2}{H^2} \geq \frac{9}{4}$).
In this case, from Eq.~(\ref{eq:app-im}) in Appendix, Hankel function for $\epsilon \ll 1$ is approximated as 
\begin{equation}
\begin{split}
|H_{i \mu}^{(1)} (\epsilon)|^2
&\simeq 
\begin{cases}
&\frac{2 \mathrm{e}^{\pi \mu}}{\pi} \left( \frac{\coth \pi \mu}{\mu} - \frac{4 \pi \mathrm{e}^{- 2 \pi \mu}}{(1 - \mathrm{e}^{- 2 \pi \mu})^2}~\mathrm{Re} \frac{\mathrm{e}^{2 i \mu \ln \frac{\epsilon}{2}}}{\Gamma (1 + i \mu)^2} \right)
~~\left( \frac{\tilde m^2}{H^2} > \frac{9}{4} \right), \\
&\frac{4}{\pi^2} \left( \ln \frac{\epsilon}{2} \right)^2~~\left( \frac{\tilde m^2}{H^2} \simeq \frac{9}{4} \right).
\end{cases}
\end{split}
\end{equation}
Then, the dimensionless noise correlation function is given by
\begin{equation}\label{eq:cor-mu-phi}
\begin{split}
&\bar S^{(\phi)} (r, N; \mathrm{d} N) \\
&= \frac{\mathrm{d} N}{(2 \pi)^2}~j_0 (\epsilon a H r)~\mathrm{e}^{- \pi \mu} \times \frac{\pi}{2} \epsilon^3~|H_{i \mu}^{(1)} (\epsilon)|^2 \\
&\simeq \frac{\mathrm{d} N}{(2 \pi)^2}~j_0 (\epsilon a H r) \times 
\begin{cases}
&\epsilon^3 \frac{\coth(\pi \mu) - 1}{\mu}~~\left( \frac{\tilde m^2}{H^2} \geq \frac{25}{4} \right), \\
&\epsilon^3 \frac{\mu}{4} ( \coth (2 \pi) - 1)+ \frac{2 \epsilon^3}{\pi} \left( \ln \frac{\epsilon}{2} \right)^2 \left( 1 - \frac{\mu}{2} \right)~~\left( \frac{9}{4} \leq \frac{\tilde m^2}{H^2} \leq \frac{25}{4} \right).
\end{cases}\\
\end{split}
\end{equation}
Notice that the zero-point noise contribution given by Eq.~(\ref{eq:zeronoise}) is removed in the third line in Eq.~(\ref{eq:cor-mu-phi}).
Even if we keep the zero-point contribution, the noise term $\bar S^{(\phi)} (r, N; \mathrm{d} N) $ (for $\epsilon \ll 1$) is still small for large effective mass $\tilde m$
since $\bar S^{(\phi)} (r, N; \mathrm{d} N)$ is almost inversely proportional to $\tilde m$.
As in the case with real $\nu$, we extrapolate $\frac{\tilde m^2}{H^2} = \frac{25}{4}$ to $\frac{\tilde m^2}{H^2} = \frac{9}{4}$ in the second line in Eq.~(\ref{eq:cor-mu-phi}) for the sake of easier analysis.
We use this extrapolated formula in the numerical calculation in Section.~\ref{sec:num}.
We have checked that the numerical results in Section.~\ref{sec:num} is insensitive to this choice of the boundary $\frac{\tilde m^2}{H^2} = \frac{25}{4}$ as long as $\epsilon \ll 1$.

In the same way, we approximate the noise correlation functions for the conjugate $\pi (x)$ ($\epsilon \ll 1$) as
\begin{equation}\label{eq:cor-mu-pi}
\begin{split}
&\bar{S}^{(\pi)} (r, N; \mathrm{d} N) \\
&= \frac{\mathrm{d} N}{(2 \pi)^2}~j_0 (\epsilon a H r) \mathrm{e}^{- \pi \mu} \times \frac{\pi}{2} ~\epsilon^3~\left| \left( \frac{3}{2} - i \mu \right) H_{i \mu}^{(1)} (\epsilon) + \epsilon H_{i \mu - 1}^{(1)} (\epsilon) \right|^2 \\
&\simeq \frac{\mathrm{d} N}{(2 \pi)^2}~j_0 (\epsilon a H r) \times 
\begin{cases}
&\epsilon^3 \mu \left( 1 + \frac{9}{4 \mu^2} \right) ( \coth (\pi \mu) - 1 ) ~~\left( \frac{\tilde m^2}{H^2} \geq \frac{25}{4} \right) \\
&\epsilon^3 \mu \frac{25}{16} ( \coth (2 \pi) - 1) + \frac{9 \epsilon^3}{2 \pi} \left( \ln \frac{\epsilon}{2} \right)^2 \left(1 -  \frac{\mu}{2} \right)~~\left( \frac{9}{4} \leq \frac{\tilde m^2}{H^2} \leq \frac{25}{4} \right).
\end{cases}
\end{split}
\end{equation}
Here, the zero-point noise contribution given by Eq.~(\ref{eq:zeronoise}) is removed in the third line.
In contrast to $\bar{S}^{(\phi)} (r, N; \mathrm{d} N)$, the behavior of $\bar{S}^{(\pi)} (r, N; \mathrm{d} N)$ is quite different for large effective mass $\tilde m$
if we keep the zero-point contribution.
This is because $\bar{S}^{(\pi)} (r, N; \mathrm{d} N) \propto \epsilon^3 \tilde m~\mathrm{e}^{- 2 \tilde m/H}$ when we remove the zero-point contribution,
while $\bar{S}^{(\pi)} (r, N; \mathrm{d} N) \propto \epsilon^3 \tilde m$ when the zero-point contribution is included for $\tilde m \gg H$.
Notice that $\bar{S}^{(\phi)} (r, N; \mathrm{d} N)$ and $\bar{S}^{(\pi)} (r, N; \mathrm{d} N)$ follow the ordinary uncertainty relation when the zero-point fluctuation contributions are included.
In the numerical analysis below, we take $\epsilon =10^{-2}$ and thus the zero-point fluctuation contribution to the noise term never become important at least for $N \lesssim 10^3$.
Now, we note that the numerical results in Section.~\ref{sec:num} is insensitive to the boundary $\frac{\tilde m^2}{H^2} = \frac{25}{4}$.
The reason for the insensitivity to the above artificial boundaries $ \frac{\tilde m^2}{H^2} = 2, \frac{25}{4}$ is that
the noise terms with $2 \leq \frac{\tilde m^2}{H^2} < \infty$ are so small (for $\epsilon \ll 1$) that these terms cannot dominate the behavior of IR mode.
In the numerical calculation, we have checked that the behavior of IR mode is determined by the noise terms only with $\frac{\tilde m^2}{H^2} \leq 2$ for $\epsilon = 10^{-2}$.

\section{Numerical analysis for flat and non-flat direction systems}
\label{sec:num}
In this section, we prepare for the numerical calculation of the Langevin equations.
The model we consider here is taken from Ref.~\cite{Enqvist:2011pt} as a concrete example.
Then, we show the numerical results and discuss about the feature of the time evolutions of IR modes.

\subsection{Setup}
Here, we rescale Eq.~(\ref{eq:Langevin}) for the purpose of the numerical calculation.
We rescale fields by Hubble scale $H$ during inflation as
\begin{equation}
\begin{split}
&\Phi \to \bar \Phi = \Phi / H,~~s^{(\phi)} \to \bar s^{(\phi)} = s^{(\phi)} / H^2, \\
&\Pi \to \bar \Pi = \Pi / H^2,~~s^{(\pi)} \to \bar s^{(\pi)} = s^{(\pi)} / H^3, \\
&V (\Phi) \to \bar V (\bar \Phi) = V (\Phi) / H^4,
\end{split}
\end{equation}
where bar shows the rescaled quantity.
Using these dimensionless quantities, Eq.~(\ref{eq:Langevin}) is rewritten as
\begin{equation}\label{eq:Lan2}
\begin{split}
& \frac{\mathrm{d}}{\mathrm{d} N} \bar \Phi (N) = \bar \Pi (N)  + \bar s^{(\phi)} (N), \\
&\frac{\mathrm{d}}{\mathrm{d} N} \bar \Pi (N) = - 3 \bar \Pi (N) - \frac{\partial \bar V (\bar \Phi (N))}{\partial \bar \Phi} + \bar s^{(\pi)} (N),
\end{split}
\end{equation}
where we have omitted the $\bold{x}$-dependence of the noise terms, since IR mode is almost spatially homogeneous inside the horizon.
We note that the time argument is now converted to the e-folding number $N = H t$.
Discretizing Eq.~(\ref{eq:Lan2}), we obtain the following dimensionless Langevin equations: 
\begin{equation}\label{eq:numlangevin}
\begin{split}
& \bar \Phi (N + \mathrm{d} N) =  \bar \Phi (N) + \bar \Pi (N) \mathrm{d} N + \bar{\xi}^{(\phi)}, \\
& \bar \Pi (N + \mathrm{d} N) =  \bar \Pi (N) - 3 \bar \Pi (N) \mathrm{d} N - \frac{\partial \bar V (\bar \Phi (N))}{\partial \bar \Phi} \mathrm{d} N + \bar{\xi}^{(\pi)}.
\end{split}
\end{equation}
Here, the noise terms $\bar{\xi}^{(\phi)}$, $\bar{\xi}^{(\pi)}$ are the Gaussian random variables satisfying the variance given by
Eqs.~(\ref{eq:noise-cor}), (\ref{eq:noise-cor-pi}), (\ref{eq:cor-mu-phi}) and (\ref{eq:cor-mu-pi}) depending on the effective mass $\tilde m$.
Although Eq.~(\ref{eq:numlangevin}) is derived for a single real scalar field case, generalization to the multi-field case is straightforward.
In the next subsection, we show the numerical results for the coupled Langevin equations with three real scalar fields,
where all noise correlation functions between different fields vanish.

\subsection{Numerical analysis}
Following Ref.~\cite{Enqvist:2011pt}, we consider the following scalar potential as a generic model for a flat direction $\phi$ and non-flat directions $\bar e$ \footnote{
Do not confuse the "bar" symbol in $\bar e$ with in quantities $\bar \Phi$ etc. . $\bar e$ is the right-handed slepton and $\bar \Phi$ etc. are the dimensionless quantities. 
}, $h$:
\begin{equation}\label{eq:potential}
\begin{split}
V
&= \frac{1}{2} \lambda_e^2 \left( \phi^2 + h^2 \right) \bar e^2 + \frac{1}{8} g_2^2 h^4 + \frac{1}{8} g_1^2  \left( h^4 + 4 \bar e^4 -4 h^2 \bar e^2 \right) + \frac{\phi^6}{M_{\text{P}}^2},
\end{split}
\end{equation}
where $\lambda_e, g_1, g_2$ are the coupling constants.
Eq.~(\ref{eq:potential}) is motivated by the $L H_u$ direction $\phi$ in minimal supersymmetric standard model (MSSM).
Here, the Hubble-induced effective masses~\cite{hep-ph/9405389, Dine:1995uk} are assumed to be absent.
This happens within D-term inflation scenarios~\cite{hep-ph/9405389} or when imposing a Heisenberg symmetry on K$\ddot {\text{a}}$hler potential~\cite{Gaillard:1995az}, for example.
In Eq.~(\ref{eq:potential}), the last term is the non-renormalizable term and $M_{\text{P}} \simeq 2.4 \times 10^{18}~\mathrm{GeV}$ is the reduced Planck scale.
For the potential Eq.~(\ref{eq:potential}), the Langevin equations are written as
\begin{equation}\label{eq:full-numlangevin}
\begin{split}    
& \bar \Phi (N + \mathrm{d} N) =  \bar \Phi (N) + \bar \Pi (N) \mathrm{d} N + \bar{\xi}^{(\phi)}, \\
& \bar \Pi (N + \mathrm{d} N) =  \bar \Pi (N) - 3 \bar \Pi (N) \mathrm{d} N 
- \left( \lambda_e^2 \bar E (N)^2 \bar \Phi (N) + 6 \frac{H^2}{M_{\text{P}}^2} \bar \Phi (N)^5 \right) \mathrm{d} N + \bar{\xi}^{(\pi_{\phi})}, \\
&\bar E (N + \mathrm{d} N) =  \bar E (N) + \bar \Pi_e (N) \mathrm{d} N + \bar{\xi}^{(e)}, \\
&\bar \Pi_e (N + \mathrm{d} N) =  \bar \Pi_e (N) - 3 \bar \Pi_e (N) \mathrm{d} N \\
&~~~~~~~~~~~~~~~~~~~~~~~~~~~~ - \left( \left( \lambda_e^2  \bar \Phi (N)^2 + \left( \lambda_e^2 - g_1^2 \right) \bar h(N)^2 \right) \bar E (N) + 2 g_1^2 \bar E (N)^3 \right) \mathrm{d} N + \bar{\xi}^{(\pi_e)}, \\
& \bar h (N + \mathrm{d} N) =  \bar h (N) + \bar \Pi_h (N) \mathrm{d} N + \bar{\xi}^{(h)}, \\
& \bar \Pi_h(N + \mathrm{d} N ) = \bar \Pi_h (N) - 3 \bar \Pi_h (N) \mathrm{d} N\\
&~~~~~~~~~~~~~~~~~~~~~~~~~~~~ - \left( \left( \lambda_e^2 - g_1^2 \right) \bar E(N)^2 \bar h(N) + \frac{1}{2} \left( g_1^2 + g_2^2 \right) \bar h(N)^3 \right)  \mathrm{d} N +\bar{\xi}^{(\pi_h)}.
\end{split}
\end{equation}
Here, $\Phi~(E, h)$ and $\Pi~(\Pi_e, \Pi_h)$ are IR modes for the scalar field $\phi~(\bar e, h)$ and its conjugate.
In the numerical analysis, we set all the fields at the origin initially:
\begin{equation}
\begin{split}
&\bar \Phi (0) = \bar E (0) = \bar h (0) = 0,\\
&\bar \Pi (0) = \bar \Pi_e (0) = \bar \Pi_h (0) = 0.
\end{split}
\end{equation}
In the following, we omit the non-renormalizable term in Eq.~(\ref{eq:full-numlangevin}) for a moment, which never becomes important within the scope of this study.

The noise terms depend on the effective masses of the scalar fields.
For the sake of convenience, we write down the effective masses:
\begin{equation}\label{eq:full-eff-mass}
\begin{split}
&\tilde m_{\phi}^2/H^2 =  \lambda_e^2 \bar E (N)^2, \\
&\tilde m_e^2/H^2 = \lambda_e^2 \bar \Phi (N)^2 + \left( \lambda_e^2 - g_1^2 \right) \bar h(N)^2 + 6 g_1^2 \bar E (N)^2, \\
&\tilde m_h^2/H^2 = \left( \lambda_e^2 - g_1^2 \right) \bar E(N)^2 + \frac{3}{2} \left( g_1^2 + g_2^2 \right) \bar h(N)^2,
\end{split}
\end{equation}
where $\tilde m_{\phi}, \tilde m_e$ and $\tilde m_h$ are the effective masses of the scalar fields $\phi, \bar e$ and $h$, respectively.
As we have mentioned below Eq.~(\ref{eq:effave}), we have replaced the spatial averages by IR mode field values in Eq.~(\ref{eq:full-eff-mass}).
In the numerical calculation, we read out the effective masses $\tilde m_{\phi}, \tilde m_e$ and $\tilde m_h$, and then determine the noise terms step by step.

In the numerical calculation, we take the time-step as $dN = 10^{-2}$.
For generating random numbers, we utilize the Mersenne Twister method~\cite{MT}.
Then, we obtain Gaussian noises by Box-Muller method with the random numbers.
The number of trials in our calculation is $5 \times 10^4$.

\subsection{The results}
\begin{figure}
\begin{center}
\includegraphics[width=100mm]{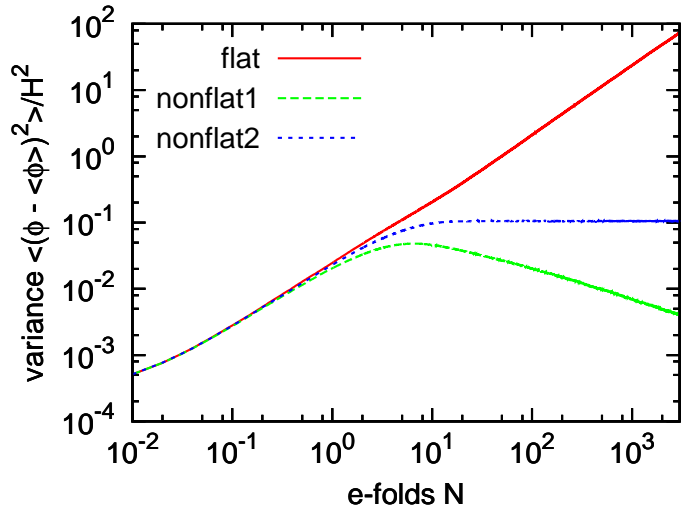}
\caption{
The time evolutions of the variances for IR modes. 
The red solid line,  the green dashed line, and the blue dotted line show the IR modes for the direction $\phi$ ("flat"), $\bar e$ ("nonflat 1"), and $h$ ("nonflat 2"), respectively.
Here, the coupling constants are $\lambda_e = g_1 = g_2 = 1.0$ and $\epsilon = 10^{-2}$.
}
\label{fig:v}
\end{center}
\end{figure}

\begin{figure}
\begin{center}
\includegraphics[width=100mm]{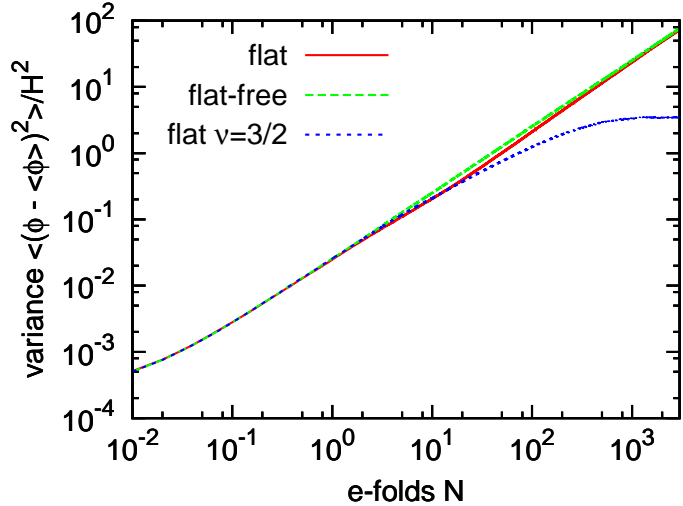}
\caption{
Comparison with flat direction IR modes for various cases.
The red solid line is the same as in Fig.~\ref{fig:v} and the green dashed line is for $\lambda_e = 0,  g_1 = g_2 = \text{arbitrary}$.
The blue dotted line is the same as Ref.~\cite{Enqvist:2011pt} but with $\epsilon = 10^{-2}$ for $\lambda_e = g_1 = g_2 = 1.0$ where the massless noise terms are used ($\nu=3/2$ for $\phi, \bar e, h$).
}
\label{fig:v4}
\end{center}
\end{figure}

Here, we show the numerical results for Eq.~(\ref{eq:full-numlangevin}).
In the numerical analysis, we take $\epsilon = 10^{-2}$ to guarantee the validity of approximation for Hanckel functions.
We have checked that the result is almost independent of $\epsilon \leq 1.0$, if we remove the zero-point fluctuation from the noise terms.
We also take the coupling constants as $\lambda_e = g_1 = g_2 = 1.0$~\footnote{
Although these coupling constants $\lambda_e = g_1 = g_2 = 1.0$ are not realistic ones in MSSM, the generic feature of the flat direction IR mode can be studied.
It is because, as we see in Fig.~\ref{fig:v4}, the flat direction reaches an exactly flat direction at late time and coupling constants affect only to the relaxation time (and saturation value for $h$ in Fig.~\ref{fig:v}).
}.
We note that, for these couplings, Eq.~(\ref{eq:full-eff-mass}) is reduced to
\begin{equation}
\begin{split}
&\tilde m_{\phi}^2/H^2 =  \lambda_e^2 \bar E (N)^2, \\
&\tilde m_e^2/H^2 = \lambda_e^2 \bar \Phi (N)^2 + 6 g_1^2 \bar E (N)^2, \\
&\tilde m_h^2/H^2 =  \frac{3}{2} \left( g_1^2 + g_2^2 \right) \bar h(N)^2.
\end{split}
\end{equation}
%

In Fig.~\ref{fig:v}, we show the time evolutions of the variances for IR modes.
The red solid line,  the green dashed line, and the blue dotted line show the IR modes for the direction $\phi$ ("flat"), $\bar e$ ("nonflat 1"), and $h$ ("nonflat 2"), respectively.
Although all IR modes' variances are degenerate until the first one Hubble time (e-folds $N=1$) pasts,
they split completely after e-folds $N \simeq 10$ since each effective mass evolves differently.
As we can see in Eq.~(\ref{eq:full-numlangevin}), non-flat direction $h$ is decoupled from others for $\lambda_e = g_1$.
Thus, the time evolution of IR mode for $h$ is determined by the self-coupling term $h^3$ and the noise terms.
From this self-coupling, the effective mass $\langle \tilde m_h^2 \rangle /H^2 = 3 \langle \bar h^2 \rangle \ll 1$ is generated.
The saturated value for $h$ ($\simeq 0.11$) in Fig.~\ref{fig:v} can be understood by the formula for the variance with $\tilde m \ll H$~\cite{Linde:1982uu}: $\langle h^2 \rangle = \frac{3 H^4}{8 \pi^2 \langle \tilde m^2 \rangle}$.
Using this formula, we obtain the averaged value $\langle \bar h^2 \rangle = \sqrt{1/(8 \pi^2)} \simeq 0.11$ which is consistent with the saturated value in  Fig.~\ref{fig:v}.
On the other hand, the flat direction $\phi$ and the non-flat direction $\bar e$ couple with each other.
Since $\bar e$ has a self-coupling term and $\phi$ does not, the effective mass for $\bar e$ become lager than the one for $\phi$.
Thus, although the variances of their IR modes are degenerate at first,
they split after the self-coupling of $\bar e$ becomes non-negligible in the effective mass.
At $N \simeq10^3$, $\bar e$ becomes too massive and cannot have a large fluctuation.
In contrast, $\phi$ eventually becomes as an exactly flat direction. See also Fig.~\ref{fig:v4} for the comparison with the case when $\phi$ is exactly free. 

\begin{figure}
\begin{center}
\includegraphics[width=75mm]{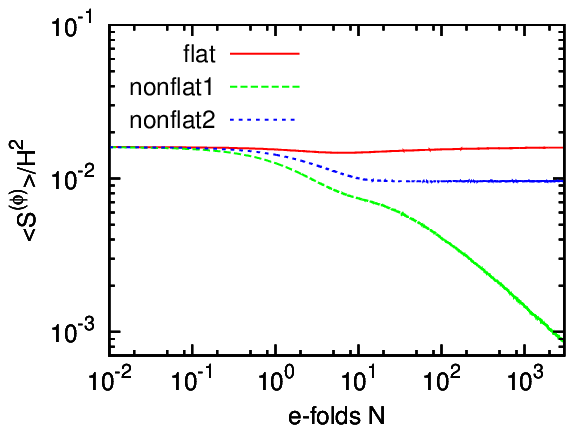}
\includegraphics[width=75mm]{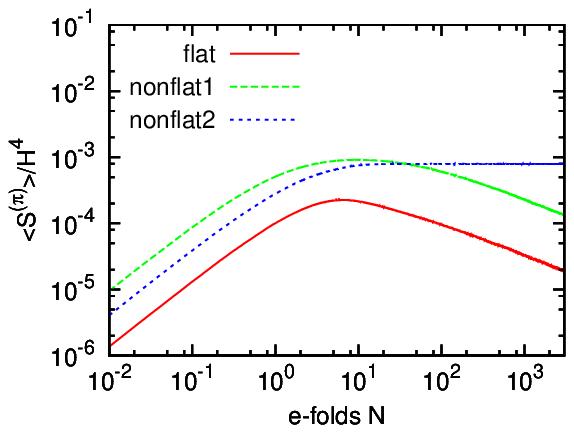}
\end{center}
\caption{
The averaged variances for the integrated noise correlation functions $\langle \bar S^{(\phi)} \rangle$ (left panel) and $\langle \bar S^{(\pi)} \rangle$ (right panel).
The noises are integrated for intervals $\mathrm{d}N = 10^{-2}$.
The lines, coupling constants and $\epsilon$ are the same as in Fig.~\ref{fig:v}.
}
\label{fig:v23}
\end{figure}
In Fig.~\ref{fig:v23}, we show the averaged variances for the integrated noise correlation functions $\langle \bar S^{(\phi)} \rangle$ (left panel) and $\langle \bar S^{(\pi)} \rangle$ (right panel).
The noises are integrated for intervals $\mathrm{d}N = 10^{-2}$.
Although the noises $\langle \bar S^{(\phi)} \rangle$ for $\phi$ and $\bar e$ are degenerate at first,
$\langle \bar S^{(\phi)} \rangle$ for $\phi$ eventually approaches the exactly flat case and $\langle \bar S^{(\phi)} \rangle$ for $\bar e$ decreases rapidly.
This is consistent with Fig.~\ref{fig:v}.
We note that $\langle \bar S^{(\pi)} \rangle$ has little effect on the numerical calculation, since $\langle \bar S^{(\phi)} \rangle  \gg \langle \bar S^{(\pi)} \rangle$.

In Fig.~\ref{fig:v4}, we show the time evolution of flat direction IR modes for various cases.
The red solid line is the same as in Fig.~\ref{fig:v} and the green dashed line is the exactly flat direction case ($\lambda_e = 0,  g_1 = g_2 = \text{arbitrary}$).
The blue dotted line is the same as Ref.~\cite{Enqvist:2011pt} but with $\epsilon = 10^{-2}$ for $\lambda_e = g_1 = g_2 = 1.0$ where the massless noise terms are used ($\nu=3/2$ for $\phi, \bar e, h$).
From Fig.~\ref{fig:v4}, we confirm that our result is much different from the one in Ref.~\cite{Enqvist:2011pt}.
This is because Ref.~\cite{Enqvist:2011pt} does not include the effective mass effects in the noise terms.
When the e-fold number is small, our result and Ref.~\cite{Enqvist:2011pt} are consistent with each other since all of the effective masses are sufficiently small.
However, after the effective mass for non-flat direction $\bar e$ becomes large $N \simeq 10$, 
the flat direction $\phi$ becomes more flat and $\bar e$ becomes more non-flat in our formulation.
Then, the flat direction $\phi$ eventually approaches to an exactly flat direction in our case.
Thus, our conclusion is different from Ref.~\cite{Enqvist:2011pt} :
even if some non-flat directions prevent a flat direction to go away from the origin,
the flatness for the flat direction eventually recovers and the variance of flat direction increases as large as the exactly flat case.

Now, we comment on the relevance of one-loop radiative corrections~\cite{Gaillard:1995az, Garbrecht:2006aw}.
Since the tree-level effective mass of the flat direction is highly suppressed at last,
one-loop radiative correction $|\tilde m_{\phi}^2| \simeq 0.01 \lambda_e^2 H^2$ suggested in Refs.~\cite{Gaillard:1995az, Garbrecht:2006aw} generally takes over the tree-level effective mass.
Thus, it is reasonable to say that the time evolution of a flat direction is determined by one-loop radiative corrections and non-renormalizable terms.
This is the same situation as one may usually have considered.
We note that the zero-point fluctuation in the noise terms has been removed in the above argument.
The case where we include the zero-point fluctuation is discussed in the next subsection.

\begin{figure}
\begin{center}
\includegraphics[width=75mm]{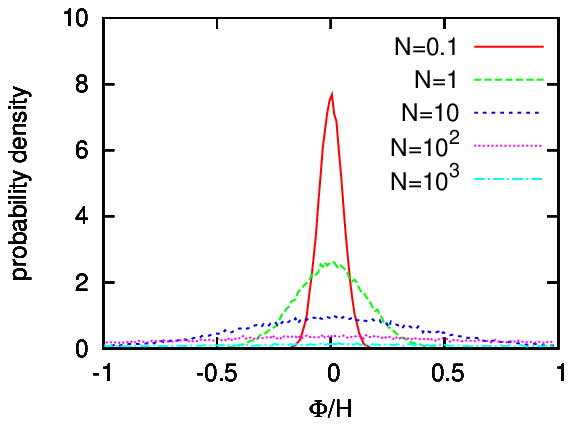}
\includegraphics[width=75mm]{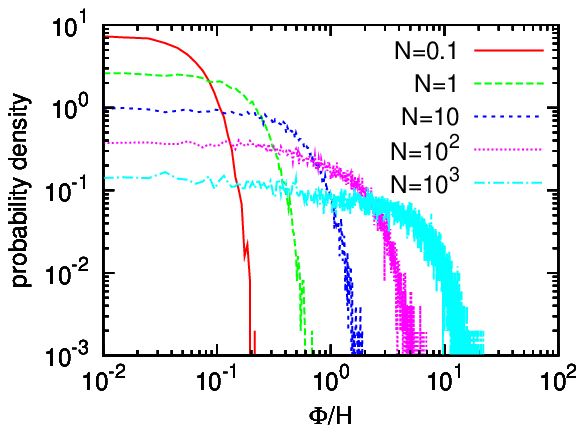}
\includegraphics[width=75mm]{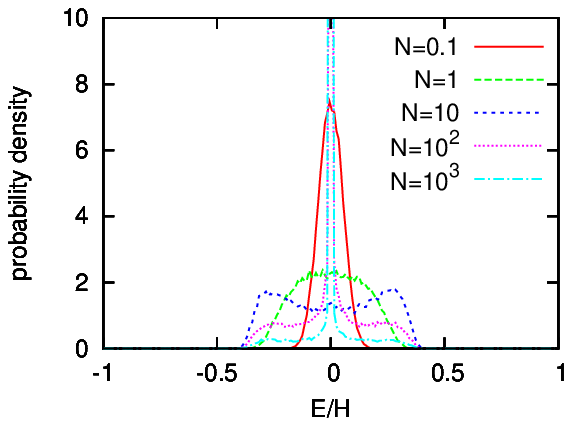}
\includegraphics[width=75mm]{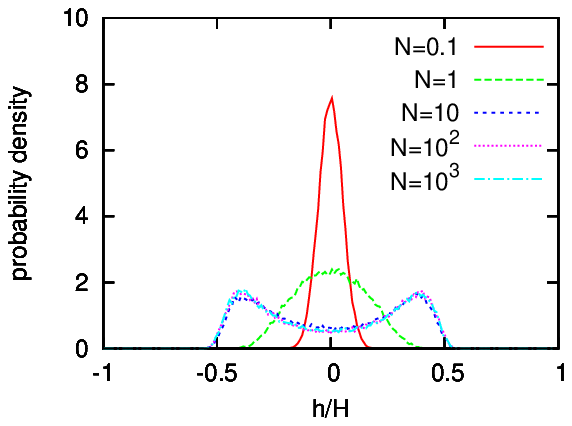}
\end{center}
\caption{
The time evolutions of the probability density functions.
The top-left, top-right, bottom-left and bottom-right panels are for flat direction $\phi$, flat direction $\phi$ (log scale), non-flat directions $\bar e$ and $h$, respectively. 
The red solid line, the green dashed line, the blue dotted line , the purple thin dotted line and the emerald dash-dotted line
are the probability density functions for $N=0.1, 1, 10, 10^2$ and $10^3$, respectively.
The coupling constants and $\epsilon$ are the same as in Fig.~\ref{fig:v}.
}
\label{fig:prob}
\end{figure}
Finally, in Fig.~\ref{fig:prob}, we show the time evolutions of the probability density functions
for the flat direction $\phi$ and the non-flat directions $\bar e$, $h$.
We construct the probability density functions by dividing the realizations into bins.  
We confirm that the flat direction $\phi$ diffuses as time goes.
As the effective mass for $\phi$ approaches to $0$ for $N \gtrsim 10$,
$\phi$ approaches to the exactly flat direction and will continue to diffuse for $N \geq 10^3$.
On the other hand, although the non-flat direction $\bar e$ diffuses at  first, $\bar e$ concentrates on the origin after $N = 10$
(the  values in Fig.~\ref{fig:prob} for $\bar e$ at $N=10^2, 10^3$ are about $20, 40$, respectively).
This is because $\bar e$ gets a large effective mass, namely,
the noise term for $\bar e$ is small and the potential enforces $\bar e$ to go back to the origin for $N \gtrsim 10$.
This is consistent with Figs.~\ref{fig:v} and \ref{fig:v23}.
The probability density function for $h$ does not change for $N \geq 10$.
This is because the variance of $h$ saturate at $N \simeq 10$.
Since the potential force is zero at the origin, the stable point is driven to non-zero $h$ by the noise terms.
This feature is also seen in the probability density function for $\bar e$ in Fig.~\ref{fig:prob} at $N = 10, 10^2$ and $10^3$.

\subsection{Discussions}
Now, let us comment on the zero-point fluctuation contribution.
If we include the zero-point contribution in the noise terms in Eq.~(\ref{eq:cor-mu-pi}),
the noise term $\bar{S}^{(\pi)} (r, N; \mathrm{d} N) \propto \epsilon^3 \tilde m$ (for $\tilde m \gg H$) eventually becomes large
and induces a non-negligible effective mass to the flat direction through the coupling $\lambda_e^2 \bar \Phi^2 \bar E^2$ after many Hubble time past. 
Then, the variance of the flat direction will saturate at last.
This means that the time evolution of the flat direction will be quite different from the case where we remove the zero-point contribution.
Fortunately, however, the existence of one-loop radiative corrections make the fate of flat direction during inflation not so different from the one without the zero-point contribution,
as long as the renormalization scale is chosen appropriately.
Anyway, we do not know which is the correct way to treat the zero-point fluctuation in the noise terms.

\section{Conclusion}
\label{sec:conc}
In this study, we have analyzed the time evolution of a flat and non-flat direction system governed by coupled Langevin equations during inflation.
We have taken into account the effective mass effects on the noise terms.
In the analysis, we have removed the zero-point fluctuation contributions from the noise terms.
As the flat direction goes away from the origin, the effective masses of non-flat directions coupled to the flat direction become large.
As a consequence, such a non-flat direction cannot have large fluctuations and cannot block the growth of the variance for the flat direction.
Thus, the tree-level effective mass of the flat direction is eventually highly suppressed.
The time evolution of the flat direction is, then, determined by one-loop radiative corrections and non-renormalizable terms as one have usually considered.
We have also discussed the case where the zero-point contributions are included in the noise terms.
In this case, the variance of the flat direction will saturate at last in the tree-level argument.
However, little is known how to treat the zero-point fluctuation in the noise terms, while it is important for a massive field with the effective mass $\gg H$. 

\section*{Acknowledgment}
T.T. is grateful to Naoya Kitajima for useful conversations.
The work of T.T. is supported in part by JSPS Research Fellowships for Young Scientists.
This work is supported by Grant-in-Aid for Scientific research from
the Ministry of Education, Science, Sports, and Culture (MEXT), Japan,
No.\ 14102004 (M.K.), No.\ 21111006 (M.K.) and also 
by World Premier International Research Center
Initiative (WPI Initiative), MEXT, Japan. 

\section*{Appendix: Approximation forms for Hankel functions}
\label{sec:app}
In this appendix, we describe the approximation formulae for Hankel functions $H_{\nu}^{(1)} (\epsilon) ~(\epsilon \ll 1)$, which we use in this study. 
When $\nu = \sqrt{\frac{9}{4} - \frac{\tilde m^2}{H^2}}$ is real ($- \infty < \frac{\tilde m^2}{H^2}  \leq \frac{9}{4}$), 
we can use the following well-known approximation formula for Hankel function~\cite{Lebedev}:
\begin{equation}\label{eq:app-wellknown}
\begin{split}
H_{\nu}^{(1)} (\epsilon)
&\simeq  \frac{\Gamma(\nu)}{i \pi} \left( \frac{\epsilon}{2} \right)^{- \nu},
\end{split}
\end{equation}
where $\epsilon \ll 1$ and $\Gamma (\nu)$ is Gamma function.
Its square is given by
\begin{equation}\label{eq:app-wellknown2}
\begin{split}
| H_{\nu}^{(1)} (\epsilon) |^2
&\simeq  \frac{2^{2 \nu} \Gamma(\nu)^2}{\pi^2} \epsilon^{- 2 \nu}.
\end{split}
\end{equation}
We cannot use Eqs.~(\ref{eq:app-wellknown}),~(\ref{eq:app-wellknown2}) for $\nu \simeq 0$ and $\nu = i \mu$.
When $\nu = 0$, Hankel function is  given by
\begin{equation}\label{eq:app-int}
\begin{split}
&H_{0}^{(1)} (\epsilon) \simeq 1 - i \frac{2}{\pi} \ln \frac{\epsilon}{2}~~(\epsilon \ll 1), \\
&|H_{0}^{(1)} (\epsilon)|^2 \simeq \frac{4}{\pi^2} \left( \ln \frac{\epsilon}{2} \right)^2~~(\epsilon \ll 1).
\end{split}
\end{equation}
On the other hand, when $\nu = i \mu~(\mu > 0)$,
we can apply the following approximation by using the expression for Bessel function with complex $\nu$~\cite{Enqvist:1987au, Lebedev}:
\begin{equation}
\begin{split}
J_{\nu} (\epsilon) 
&= \sum_{k=0}^{\infty} \frac{(-1)^k}{\Gamma(k+1) \Gamma(\nu + k + 1)} \left( \frac{\epsilon}{2} \right)^{\nu + 2 k} \\
&\simeq \frac{1}{\Gamma(1 + \nu)}  \left( \frac{\epsilon}{2} \right)^{\nu}~~(\epsilon \ll 1).
\end{split}
\end{equation}
Using this expression, we find the following approximation formula for Hankel function:
\begin{equation}
\begin{split}
H_{\nu}^{(1)} (\epsilon)
&= J_{\nu} (\epsilon) + i N_{\nu}^{(1)} (\epsilon) \\
&= \frac{2}{1 - \mathrm{e}^{2 i \pi \nu}} \left( J_{\nu} (\epsilon) - \mathrm{e}^{i \pi \nu} J_{- \nu} (\epsilon) \right) \\
&\simeq \frac{2}{1 - \mathrm{e}^{2 i \pi \nu}} \left( \frac{1}{\Gamma (1 + \nu)} \left( \frac{\epsilon}{2} \right)^{\nu} - \mathrm{e}^{i \pi \nu} \frac{1}{\Gamma (1 - \nu)}  \left( \frac{\epsilon}{2} \right)^{- \nu} \right)~~(\epsilon \ll 1).
\end{split}
\end{equation}
Then, we find for $\nu = i \mu~(\mu >0)$:
\begin{equation}\label{eq:app-im}
\begin{split}
|H_{i \mu}^{(1)} (\epsilon)|^2 
&\simeq \frac{4}{|1 - \mathrm{e}^{- 2 \pi \mu}|^2} \left( \frac{1}{|\Gamma (1 + i \mu)|^2} 
+ \frac{\mathrm{e}^{- 2 \pi \mu} }{|\Gamma (1 - i \mu)|^2}
- 2 \mathrm{e}^{- \pi \mu}~\mathrm{Re}~\frac{\mathrm{e}^{2 i \mu \ln \frac{\epsilon}{2}}}{\Gamma (1+ i \mu)^2} \right) \\
&= \frac{2 \mathrm{e}^{\pi \mu}}{\pi} \left( \frac{\coth \pi \mu}{\mu} - \frac{4 \pi \mathrm{e}^{- 2 \pi \mu}}{(1 - \mathrm{e}^{- 2 \pi \mu})^2}~\mathrm{Re} \frac{\mathrm{e}^{2 i \mu \ln \frac{\epsilon}{2}}}{\Gamma (1 + i \mu)^2} \right).
\end{split}
\end{equation}
%

{}

\end{document}